\documentclass[iop,apj]{emulateapj}

\usepackage{apjfonts}
\usepackage{amsmath, amstext}
\usepackage{url}
\usepackage[backref, breaklinks, colorlinks=true, urlcolor=blue, anchorcolor=blue, linkcolor=blue, citecolor=blue,]{hyperref}
\usepackage[all]{hypcap}     
\usepackage{bm}
\usepackage[normalem]{ulem}
\usepackage{color}

\slugcomment{Draft version \today}

\shorttitle{EDGES High-Band receiver calibration}
\shortauthors{Monsalve et al.}

\begin{document}

\title{Calibration of the EDGES High-Band Receiver to Observe the Global 21-cm Signature\\ from the Epoch of Reionization}

\author{
Raul A. Monsalve\altaffilmark{1,2}, 
Alan E. E. Rogers\altaffilmark{3}, 
Judd D. Bowman\altaffilmark{2}, 
and Thomas J. Mozdzen\altaffilmark{2}
}

\affil{$^1$Center for Astrophysics and Space Astronomy, University of Colorado, Boulder, CO 80309, USA; \href{mailto:Raul.Monsalve@colorado.edu}{Raul.Monsalve@colorado.edu}}
\affil{$^2$School of Earth and Space Exploration, Arizona State University, Tempe, AZ 85287, USA}
\affil{$^3$Haystack Observatory, Massachusetts Institute of Technology, Westford, MA 01886, USA}

\begin{abstract}

The EDGES High-Band experiment aims to detect the sky-average brightness temperature of the $21$-cm signal from the Epoch of Reionization (EoR) in the redshift range $14.8 \gtrsim z \gtrsim 6.5$. To probe this redshifted signal, EDGES High-Band conducts single-antenna measurements in the frequency range $90-190$ MHz from the Murchison Radio-astronomy Observatory in Western Australia. In this paper, we describe the current strategy for calibration of the EDGES High-Band receiver and report calibration results for the instrument used in the $2015-2016$ observational campaign. We propagate uncertainties in the receiver calibration measurements to the antenna temperature using a Monte Carlo approach. We define a performance objective of $1$~mK residual RMS after modeling foreground subtraction from a fiducial temperature spectrum using a five-term polynomial. Most of the calibration uncertainties yield residuals of $1$~mK or less at $95\%$ confidence. However, current uncertainties in the antenna and receiver reflection coefficients can lead to residuals of up to $20$ mK even in low-foreground sky regions. These dominant residuals could be reduced by 1) improving the accuracy in reflection measurements, especially their phase 2) improving the impedance match at the antenna-receiver interface, and 3) decreasing the changes with frequency of the antenna reflection phase.

\end{abstract}


\section{Introduction}

The sky-average, or global, component of the redshifted $21$-cm signal represents a direct tracer of the bulk characteristics of the intergalactic medium (IGM) during cosmic dawn and the epoch of reionization (EoR) at redshidfts $z \gtrsim 6$ \citep{madau1997, shaver1999, zaldarriaga2004, furlanetto2006}. The observable quantity corresponds to a differential brightness temperature that depends on the fraction of neutral hydrogen and the spin temperature of the gas. This temperature encodes the effects on the IGM of UV and X-ray radiation from the first generations of stars and stellar remnants \citep{fialkov2014a, fialkov2014b, fialkov2016a, mirocha2013, mirocha2015}. For $z \gtrsim 6$, the large-scale evolution of the IGM is captured as wideband features in the frequency spectrum of the brightness temperature below $\sim 200$ MHz, with expected absolute amplitudes lower than $\sim 200$ mK \citep{mesinger2013, fialkov2016b, mirocha2016, cohen2016}.

The most direct instrumental approach to attempt the global measurement is the single-antenna, wideband-spectrometer design implemented by EDGES \citep{bowman2008, rogers2008, bowman2010, rogers2015, mozdzen2016a, mozdzen2016b}, BIGHORNS \citep{sokolowski2015a, sokolowski2015b}, SARAS \citep{patra2013, patra2015}, and SCI-HI \citep{voytek2014}. These ground-based experiments cover differing portions of the redshift domain, targeting specific features of the global $21$-cm signal. Using a similar approach, the DARE experiment aims to conduct this measurement from the far side of the Moon to mitigate the impact of radio-frequency interference and the Earth's ionosphere \citep{burns2012, jones2015, datta2016}. The LEDA experiment also pursues the global signal through single-antenna total-power measurements, with the addition of interferometric measurements to help estimate instrumental and foreground parameters \citep{greenhill2012, bernardi2015, bernardi2016}.

EDGES has conducted sky measurements from Western Australia since 2006. Until 2012 the experiment relied on relative calibration provided by three-position switching at the input of the receiver. This approach did not fully account for the impedance mismatch between the antenna and the receiver, since reflections from the receiver input were assumed to be zero, and the system gain and antenna reflections were not referenced to the same calibration plane. In addition, the calibration did not consider a correction for beam chromaticity. Despite this simplified approach, EDGES was able to place initial constraints on the duration of the EoR and obtain a first-order estimate for the spectral index of diffuse emission in low-foreground regions \citep{bowman2008, rogers2008, bowman2010}.

In 2013, EDGES deployed the first instrument that implemented the current end-to-end absolute calibration. This approach consists of 1) converting the noise power measured by the antenna at the receiver input to an absolute antenna temperature scale, properly accounting for impedance mismatches between the receiver and the antenna, 2) removing the effect of antenna losses, and 3) compensating for the effect of beam chromaticity. Although in its first iteration it did not achieve the sensitivity for a cosmological detection, its high performance enabled quantification of perturbations in the ionosphere \citep{rogers2015}.

Starting in 2015, EDGES has operated a low-band ($50-100$ MHz) and a high-band ($90-190$ MHz) instrument. Both share the same design and absolute calibration approach, and target redshifts that nominally correspond to the cosmic dawn ($27.4\gtrsim z \gtrsim 13.2$) and EoR ($14.8 \gtrsim z \gtrsim 6.5$) periods, respectively. They represent a significant upgrade with respect to previous iterations, with the objective of reducing systematic effects below the cosmological signal. In this revision, the fourpoint antenna \citep{suh2003} was replaced with a blade model due to its lower beam chromaticity \citep{mozdzen2016a}. The antenna reflection is periodically measured in situ to high accuracy. The receiver is calibrated in the laboratory to higher precision and accuracy than in 2013, and in the field it operates underground with active temperature control. These and other upgrades have recently enabled improved long-term ($>200$ days) stable measurements of the spectral index of diffuse foregrounds \citep{mozdzen2016b}.

\begin{figure}[t]
\centering
\includegraphics[width=0.48\textwidth]{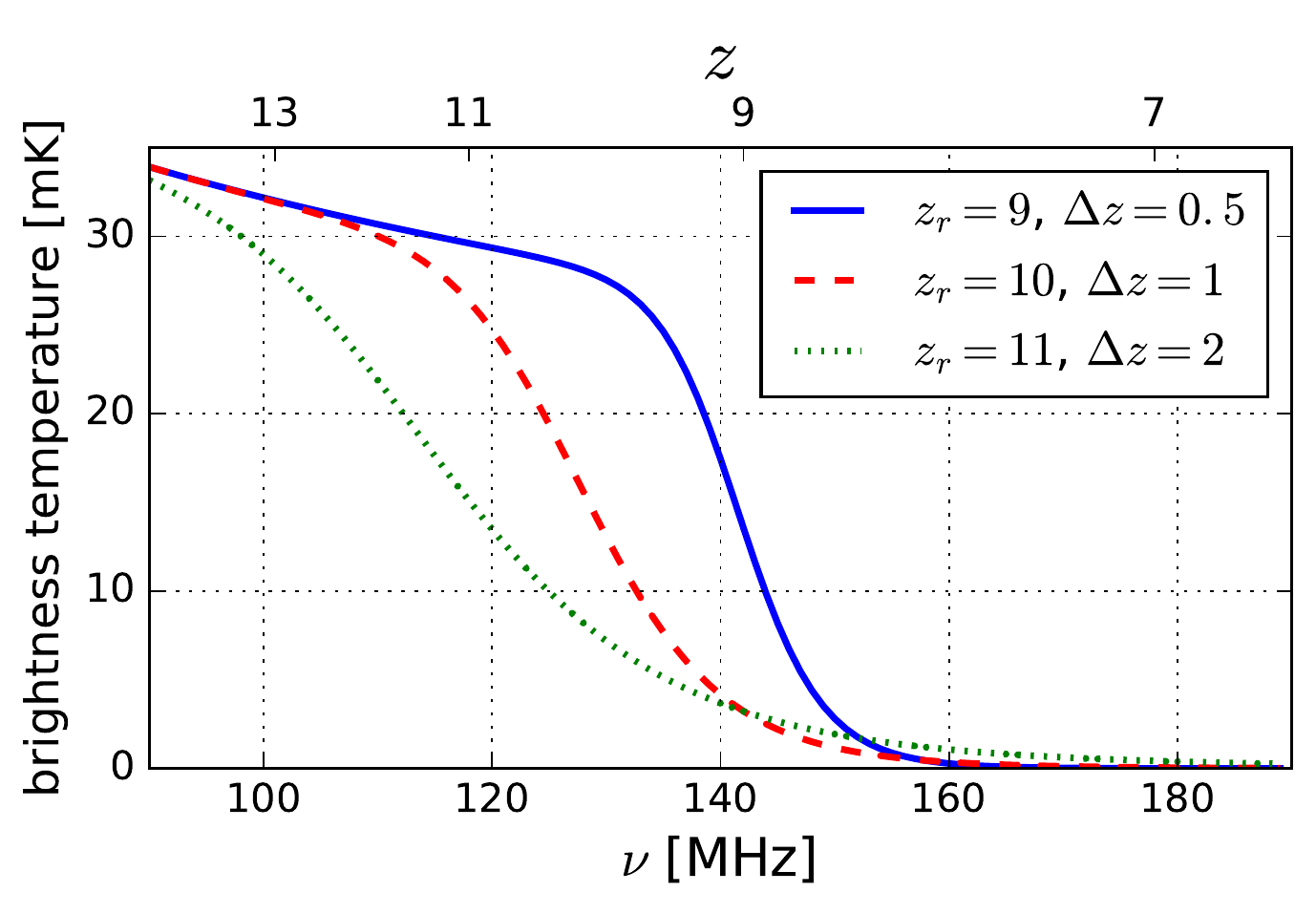}
\caption{Three reference \emph{tanh}-based models for the EoR signal, parameterized in terms of the redshift ($z_r$) and duration ($\Delta z$) of reionization.}
\label{figure_models_eor}
\end{figure}

This paper describes the laboratory calibration of the $2015-2016$ EDGES High-Band receiver, which nominally targets the EoR signal. For reference, Figure~(\ref{figure_models_eor}) shows three phenomenological models for this signal, parameterized in terms of the redshift ($z_r$) and duration ($\Delta z$) of reionization \citep{pritchard2010, bowman2010, morandi2012, liu2013, mirocha2015, harker2016}.

Our receiver calibration follows the method introduced in \cite{rogers2012}, which involves measuring the spectra, reflection coefficients, and physical temperatures of four absolute calibrators connected externally to the receiver input, in place of the antenna. These calibration measurements are used to determine the function that converts the noise power measured by the instrument to antenna temperature referenced to the receiver input. We use this function to calibrate the sky measurements obtained in the field with the receiver operating at the same temperature as in the lab. 

As part of the description of our calibration, we expand on the methodology presented in \cite{rogers2012} by providing a model for the ambient and hot calibrator used as the absolute temperature reference. We also introduce two new frequency-dependent parameters --- a scale and an offset --- that relate the relative calibration obtained through the internal switching to the absolute calibration at the receiver input.

After determining the fiducial receiver calibration, we model the uncertainties encountered in the laboratory measurements and propagate them to the calibrated antenna temperature using a Monte Carlo approach. We use simulated sky measurements as inputs in this process, and quantify the impact of potential calibration errors via the number of polynomial terms required to fit the corrupted spectrum and reduce the residuals below a nominal threshold of $1$ mK. Robust estimates are obtained for the impact of each source of uncertainty, as well as for their combined effect. In order to focus on receiver uncertainties, these simulations assume perfect removal of antenna losses and beam chromaticity.

The paper is organized as follows: Section~\ref{section_edges} describes the EDGES instrument and summarizes the receiver calibration strategy and nomenclature, Section~\ref{section_calibration_measurements} describes the calibration measurements and results, Section~\ref{section_uncertainty_propagation} describes the propagation of receiver uncertainty to the antenna temperature using Monte Carlo simulations, Section~\ref{section_results} presents the results of the uncertainty propagation, Section~\ref{section_discussion} discusses the limiting factors in the calibration performance and some alternatives for improvement, and Section~\ref{section_conclusion} summarizes the findings of this work.

\section{EDGES High-Band Instrument}
\label{section_edges}

\subsection{Description}
\label{section_instrument_description}

\begin{figure}[t]
\centering
\includegraphics[width=0.48\textwidth]{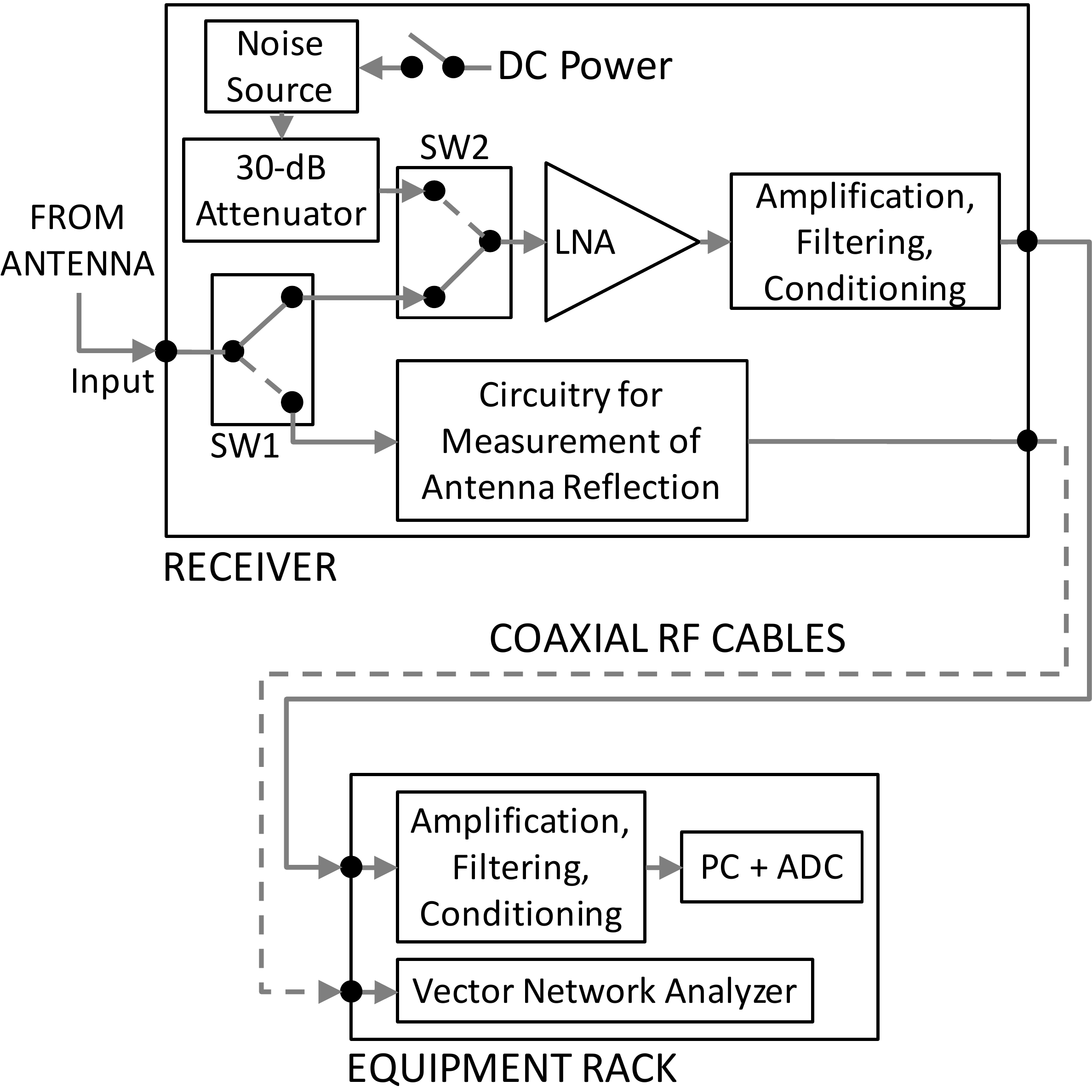}
\caption{Conceptual block diagram of the EDGES instrument.}
\label{figure_block_diagram}
\end{figure}

Figure~\ref{figure_block_diagram} presents a conceptual block diagram of the EDGES instrument. The receiver that operates in the field corresponds to the actual unit calibrated in the lab and it is part of an identical setup, except for the antenna and ground plane which only exist in the field.

Physically, the receiver consists of a metal enclosure that houses the first-stage low-noise amplifier (LNA), noise references for relative calibration, additional stages of amplification, filtering, and conditioning, and electronics for remote measurement of the antenna reflection coefficient. The receiver operates at $25^{\circ}$C at all times in order to keep the noise and reflection characteristics of the LNA stable during calibration and sky observations. Temperature stability to better than $0.1^{\circ}$C is achieved using a thermal controller that reads out a thermistor mounted inside the receiver. Based on the reading, the controller acts on a hot/cold plate attached to the bottom of the receiver to compensate for temperature drifts.

From the output of the receiver the RF signal is taken along a coaxial cable into an equipment rack where it is amplified and filtered by back-end electronics. The signal is then sent to a PC-based $400$-MS/s analog-to-digital converter (ADC) that samples with $14$-bit resolution. Finally, blocks of $65536$ time samples are Fourier transformed to the frequency domain to obtain spectra with $32768$ points and $6.1$-kHz resolution using a four-term Blackman-Harris window function.

The LNA input switches continuously between the antenna and an internal calibration position using switch SW2 in Figure~\ref{figure_block_diagram}. When in the calibration position, the LNA receives sequentially two noise levels: 1) load and 2) load + noise source. In the load state, the noise comes from the output of a $30$-dB attenuator, and in the load + noise source state the active noise source connected to the input of the attenuator is turned on for additional noise. 

The noise from the antenna only represents a few percent of the total power at the receiver output. Most of the output power is due to out-of-band noise injected below $30$ MHz as part of the conditioning stage. This injected noise remains constant during the input switching and provides a stable signal level to the back-end electronics and digitizer, improving linearity and dynamic range in the measurements.    

Each of the three input levels are measured for $13$ seconds, which corresponds to an antenna duty cycle of $33.3$\%. A total of $40960$ spectra are accumulated for each position. An implicit assumption of this switching scheme is that the amplification chain remains stable within each $39$-second cycle.

The circuitry for measurement of antenna reflections, shown in Figure~\ref{figure_block_diagram}, functions as a remote calibration unit for a vector network analyzer (VNA).  It is centered around a four-position mechanical RF switch where three of the ports are connected to open, short, and matched calibration standards, and the fourth port serves as a pass-through to the antenna.

The efficiency of the instrument is highest below $195$ MHz. Therefore, the calibration described in the rest of this paper was conducted in the range $90-190$ MHz.

\subsection{Calibration Formalism}
\label{section_calibration_formalism}

Here we summarize the calibration strategy used by EDGES, which is based on the method introduced in \citet{rogers2012}. 

For each three-position cycle of the receiver, the power spectral density (PSD) from the antenna ($P_{\text{ant}}$), load ($P_{\text{L}}$), and load + noise source ($P_{\text{L+NS}}$), are used to compute an initial \emph{uncalibrated} antenna temperature,

\begin{equation}
T^*_{\text{ant}} = T_{\text{NS}}  \frac{\left(P_{\text{ant}} - P_\text{L}\right)}{\left(P_{\text{L+NS}} - P_\text{L}\right)} + T_\text{L},
\label{equation_uncalibrated_temperature}
\end{equation}

\noindent where $T_\text{L}$ and $T_{\text{NS}}$ represent realistic assumptions for the noise temperatures of the load and noise source, respectively. This computation serves to calibrate out the time-dependent system gain, which includes the complex bandpass of the filters, amplifiers, cables, and ADC. All the parameters in Equation~(\ref{equation_uncalibrated_temperature}), and in what follows, are frequency-dependent. We do not explicitly show this dependence for simplicity of notation.

To derive the expression for calibration of $T^*_{\text{ant}}$ it is necessary to write the PSDs in Equation~(\ref{equation_uncalibrated_temperature}) in terms of the specific instrument response contributions. The PSD for the antenna is given by:

\begin{align}
P_{\text{ant}} =\;&g\big[T_{\text{ant}}\left(1-|\Gamma_{\text{ant}}|^2\right)|F|^2 \nonumber\\
&+ T_{\text{unc}}|\Gamma_{\text{ant}}|^2|F|^2 \nonumber\\
&+ T_{\cos}|\Gamma_{\text{ant}}||F|\cos\alpha \nonumber\\
&+ T_{\sin}|\Gamma_{\text{ant}}||F|\sin\alpha \nonumber\\
&+ T_0\big],
\label{equation_power_antenna}
\end{align}

\noindent with

\begin{align}
F      =& \frac{\sqrt{1-|\Gamma_{\text{rec}}|^2}}{1-\Gamma_{\text{ant}}\Gamma_{\text{rec}}},\label{equation_F}\\
\alpha =& \arg\left(\Gamma_{\text{ant}}F\right).\label{equation_alpha}
\end{align}

Here, $T_{\text{ant}}$ corresponds to the calibrated antenna temperature. The quantities $g$ and $T_0$ represent the system gain referenced to the receiver input, and the receiver noise offset, respectively.  $\Gamma_{\text{ant}}$ is the reflection coefficient of the antenna and $\Gamma_{\text{rec}}$ is the reflection coefficient of the receiver, both referenced to a $50$-$\Omega$ system impedance. The temperatures $T_{\text{unc}}$, $T_{\cos}$, and $T_{\sin}$ are called noise wave parameters following the formalism introduced by \citet{meys1978}. They are associated with the noise emitted by the LNA input toward the antenna. This noise is reflected back due to imperfect impedance match and re-enters the receiver with phase $\alpha$. The $T_{\text{unc}}$ temperature represents the portion of input noise that is uncorrelated with the noise at the LNA output, while $T_{\cos}$ and $T_{\sin}$ are components of the correlated portion.

The PSDs for the internal load and load + noise source follow the same form as Equation~(\ref{equation_power_antenna}). However, at this stage we assume that the reflection coefficients of the load and noise source are zero. In reality, they are not zero but very low ($<-40$ dB). Thus, these PSDs are modeled as:

\begin{align}
P_{\text{L}}    &= g^*\left[T_{\text{L}}\left(1-|\Gamma_{\text{rec}}|^2\right) + T_0^*\right],\label{equation_power_internal_calibrators1}\\
P_{\text{L+NS}} &= g^*\left[\left(T_{\text{L}} + T_{\text{NS}}\right)\left(1-|\Gamma_{\text{rec}}|^2\right) + T_0^*\right].\label{equation_power_internal_calibrators2}
\end{align}

In these equations, the system gain ($g^*$) and noise offset ($T_0^*$) are not exactly the same as in Equation~(\ref{equation_power_antenna}) because the noise from the internal references is injected at SW2 instead of at the receiver input itself. This aspect, in addition to the assumption for the reflection coefficients, is accounted for below with the introduction of two new parameters fitted during calibration that are able to absorb these small effects.

Plugging the three PSD definitions into Equation~(\ref{equation_uncalibrated_temperature}) results in the following identity:

\begin{align}
\left(T^*_{\text{ant}} - T_{\text{L}}\right)C_1 + \left(T_{\text{L}} - C_2\right) & =\nonumber\\
& T_{\text{ant}}\left[\frac{\left(1-|\Gamma_{\text{ant}}|^2\right)|F|^2}{\left(1-|\Gamma_{\text{rec}}|^2\right)}\right] \nonumber\\
+\;& T_{\text{unc}}\left[\frac{|\Gamma_{\text{ant}}|^2|F|^2}{\left(1-|\Gamma_{\text{rec}}|^2\right)}\right] \nonumber\\
+\;& T_{\cos}\left[\frac{|\Gamma_{\text{ant}}||F|}{\left(1-|\Gamma_{\text{rec}}|^2\right)}\cos\alpha\right] \nonumber\\
+\;& T_{\sin}\left[\frac{|\Gamma_{\text{ant}}||F|}{\left(1-|\Gamma_{\text{rec}}|^2\right)}\sin\alpha\right].
\label{equation_main_identity}
\end{align}

This equation establishes the relationship between the uncalibrated antenna temperature $T^*_{\text{ant}}$, computed with Equation~(\ref{equation_uncalibrated_temperature}), and the calibrated antenna temperature $T_{\text{ant}}$ at the receiver  input. More generically, $T_{\text{ant}}$ represents the calibrated noise temperature of any device under measurement at the receiver input. Any losses in the device under measurement have to be corrected for externally.

The quantities $C_1$ and $C_2$ introduced on the left-hand side of Equation~(\ref{equation_main_identity}) represent a scale and an offset that correct the first-order assumptions used for $T_{\text{L}}$ and $T_{\text{NS}}$ in Equation~(\ref{equation_uncalibrated_temperature}). They also account for the small path difference between the internal calibration position of SW2 and the receiver input. Finally, they also account to first order for the non-zero reflection coefficient of the internal load and noise source. Unaccounted higher-order effects are expected to be negligible due to the low reflection coefficients of these devices.

The reflection coefficient of the receiver input and the antenna are measured directly with a VNA. $F$ and $\alpha$ (Equations (\ref{equation_F}) and (\ref{equation_alpha})) are computed from these coefficients. Therefore, the remaining calibration task consists of estimating the scale $C_1$, offset $C_2$, and the noise wave parameters $T_{\text{unc}}$, $T_{\cos}$, and $T_{\sin}$, in order to apply Equation~(\ref{equation_main_identity}). To solve for these five frequency-dependent quantities we conduct laboratory measurements of four absolute calibration standards connected to the input of the receiver in place of the antenna.

Following \citet{rogers2012}, the four calibrators are: 1) an ambient load, 2) a hot load, 3) a long ($\approx$ $8$-m) open-ended coaxial cable, and 4) the same long cable but short-circuited at its far end. The ambient and hot loads provide the main temperature references, while the open and shorted cable produces ripples in its spectra that manifest the noise properties of the receiver and enable the estimation of the noise wave parameters.  The specific quantities that need to be measured for receiver calibration are:

\begin{enumerate}
\item The uncalibrated temperature spectra for each calibrator $\big($$T^*_A$, $T^*_H$, $T^*_O$, $T^*_S$$\big)$ via Equation~(\ref{equation_uncalibrated_temperature}), 
\item The physical or noise temperature of the calibrators $\big($$T_A$, $T_{H}$, $T_O$, $T_S$$\big)$, and 
\item The reflection coefficient of the calibrators $\big($$\Gamma_A$, $\Gamma_H$, $\Gamma_O$, $\Gamma_S$$\big)$. 
\end{enumerate}
The subscripts $A$, $H$, $O$, and $S$ correspond to the ambient load, hot load, open cable, and shorted cable, respectively.

\section{Calibration Measurements}
\label{section_calibration_measurements}

\begin{figure}[t]
\centering
\includegraphics[width=0.48\textwidth]{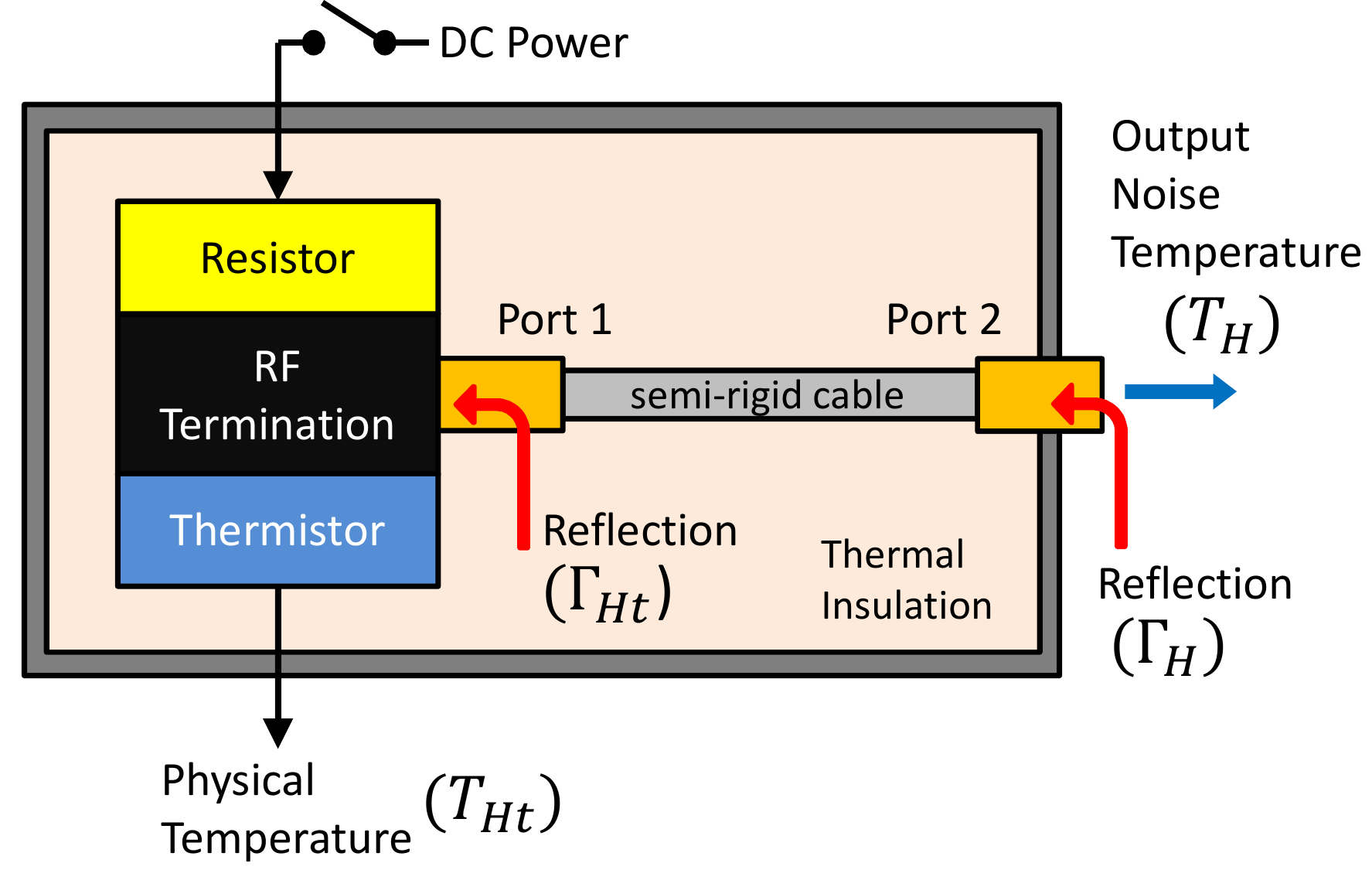}
\caption{Diagram of device used as ambient and hot load for calibration of the EDGES receiver. The noise at the output of the device depends on the physical temperatures of the RF termination and the semi-rigid cable, as well as on the reflections between blocks and the S-parameters of the cable (see equations \ref{equation_cable_loss} and \ref{equation_available_power_gain}). The nomenclature given in the figure corresponds to the device operating as a hot load.}
\label{figure_hot_load_device}
\end{figure}

In this section we report details of the laboratory measurements.  We begin by describing the calibrator used as ambient and hot load. We then discuss the measurements of spectra, physical temperature, and reflection coefficient of the calibrators. We also introduce the main sources of uncertainty for each calibration measurement, but defer detailed discussion to Section~\ref{section_uncertainty_propagation}. We conclude the section with the calculation and modeling of the five frequency-dependent receiver calibration quantities.

\subsection{Absolute Ambient/Hot Load}
\label{section_absolute_hot_load}

The ambient and hot loads used for receiver calibration are implemented as a single device, depicted in Figure~\ref{figure_hot_load_device}. It consists of an RF termination of $50$-$\Omega$ nominal impedance connected to an $8$-cm semi-rigid cable inside a thermally insulated metal enclosure. When acting as an ambient load the device operates at room temperature ($\approx$ $296$~K) and when used as a hot load the termination is heated up to $\approx$ $400$~K by powering a resistor that is thermally connected to the termination.

The effective noise temperature of the device when operating as a hot load ($T_H$) is related to the physical temperature of the termination ($T_{Ht}$) and the physical temperature of the cable ($T_{\text{cab}}$) by

\begin{equation}
T_H = G T_{Ht} + \left(1-G\right) T_{\text{cab}},
\label{equation_cable_loss}
\end{equation}

\noindent where $G$ is the available power gain of the assembly, defined as \citep{pozar2004}:

\begin{equation}
G = \frac{|S_{21}|^2\left(1-|\Gamma_{Ht}|^2\right)}{|1-S_{11}\Gamma_{Ht}|^2\left(1-|\Gamma_H|^2\right)}.
\label{equation_available_power_gain}
\end{equation}

In this equation, $S_{11}$ and $S_{21}$ are two of the S-parameters of the cable, with port~1 attached to the RF termination and port~2 corresponding to the output connector of the device. $\Gamma_{Ht}$ is the reflection coefficient of the termination alone and $\Gamma_H$ is the reflection coefficient of the device as a whole.

\subsection{Measurements and Models}
The uncalibrated temperature spectra of the absolute calibrators are measured by connecting each calibrator to the receiver input in place of the antenna. These measurements involve the same internal three-position switching as with the antenna, where in each $39$-second cycle the setup measures the PSD of the calibrator and the two internal noise references. After the measurements, Equation~(\ref{equation_uncalibrated_temperature}) is applied offline to produce the uncalibrated temperature spectra. The noise temperatures assumed for the internal load and noise source at this step are $T_{\text{L}}=300$~K and $T_{\text{NS}}=350$~K, constant across frequency. The final uncalibrated spectra, $T^*_A$, $T^*_H$, $T^*_O$, and $T^*_S$, are obtained by averaging $24$ hours of data from each calibrator. This is done in order to reduce thermal noise. Residual noise is the main source of uncertainty in these quantities.

The physical temperatures, $T_A$, $T_{Ht}$, $T_O$, and $T_S$, are measured using thermistors attached to the termination of the ambient/hot load and to the long cable. These measurements are conducted in parallel to the spectra measurement of each calibrator and with a similar time resolution. They are averaged in time to have a direct correspondence with the spectra averages. Uncertainties in these measurements are dominated by potential inaccuracies in the resistance-to-temperature model used for the thermistors and unaccounted thermal gradients in the calibrators.

The reflection coefficients and S-parameters required in the calibration are measured with a VNA. In each measurement, hundreds of traces are averaged to reduce the noise to levels so that they are an insignificant source of uncertainty. The calibrators are passive devices and, thus, they are measured using a typical VNA power of $0$ dBm ($1$ mW). In contrast, the receiver input has to be measured at $-30$ dBm to avoid saturating the active electronics designed for low-level signals. Lower VNA power results in higher measurement noise, but we compensated through averaging. We calibrate the VNA immediately before every measurement and, thus, do not assume long-term VNA stability. The main uncertainty associated with these measurements arises from imperfect VNA calibration.

The isolation between the inputs of each of the mechanical switches (SW1 and SW2 in Figure~\ref{figure_block_diagram}) is about $80$ dB. However, we assume that the isolation is perfect, which has a negligible impact on the calibration. The switching repeatability is assessed through high-precision S-parameter measurements and the scatter is found to be within the measurement uncertainties.

We model and fit the uncalibrated temperature spectra, the reflection coefficients, and the S-parameters, to avoid propagating measurement noise to the fiducial calibration quantities and to bring all the measurements to the same frequency resolution. Due to the smooth frequency behavior of the ambient and hot load spectra, and of the S-parameters of the semi-rigid cable, we model these measurements as polynomials in frequency. On the other hand, we use Fourier series in frequency to model the spectra of the open and shorted cable, as well as all the reflection coefficients. Fourier series are more efficient than polynomials at capturing the higher frequency structure encountered in these measurements. The model parameters are computed through least squares using QR decomposition for better numerical stability. The fit residuals for the spectra are noise-like. They are accounted for as a source of uncertainty in the MC analysis of Section~\ref{section_uncertainty_propagation}. For the reflection coefficients and S-parameters, the RMS residuals are $<0.001$ dB in magnitude and $<0.008^{\circ}$ in phase. Their effects are negligible compared to that from the uncertainty in VNA calibration, addressed in Section~\ref{section_uncertainty_propagation}.

\subsection{Derived Calibration Quantities}
\label{section_fiducial_calibration_quantities}

\begin{figure}[t]
\centering
\includegraphics[width=0.48\textwidth]{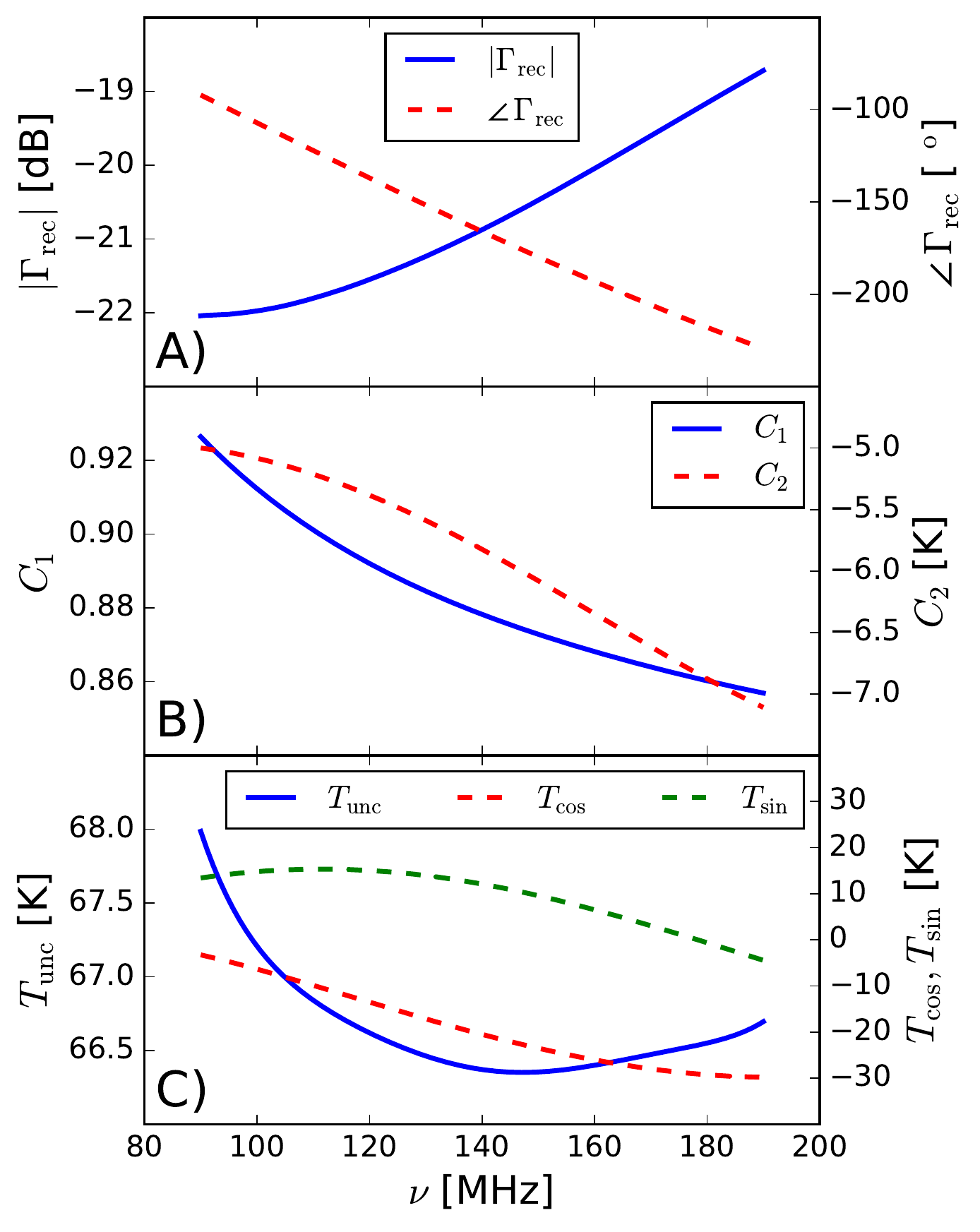}
\caption{A) The reflection coefficient of the receiver input, shown in magnitude and phase, is obtained from model fits to the VNA measurement. B) and C) Derived receiver calibration quantities. $C_1$: scale, $C_2$: offset, $T_{\text{unc}}$: uncorrelated noise wave, and $T_{\text{cos}}$, $T_{\text{sin}}$: components of the correlated noise wave. They satisfy Equation~(\ref{equation_main_identity}) when it is evaluated using the lab measurements of the absolute calibrators. All the quantities in this figure are used as the fiducial quantities in the uncertainty propagation analysis discussed in Section~\ref{section_uncertainty_propagation}.}
\label{figure_calibration_quantities}
\end{figure}

The frequency-dependent calibration quantities ($C_1$, $C_2$, $T_{\text{unc}}$, $T_{\cos}$, and $T_{\sin}$) are computed from the modeled laboratory measurements through an iterative process. Specifically, in each iteration the scale and offset are computed as

\begin{align}
C^i_1 =&\;\; C^{i-1}_1\cdot\frac{(T_H-T_A)}{(T^i_H-T^i_A)},\\ 
C^i_2 =&\;\; C^{i-1}_2 + T^i_A-T_A.  
\end{align}

Here, $T_H$ is the noise temperature of the hot load from Equation~(\ref{equation_cable_loss}), $T_A$ is the physical temperature of the ambient load, and $T^i_A$ are $T^i_H$ are the calibrated temperature spectra of the ambient and hot loads from Equation~(\ref{equation_main_identity}) evaluated at the $i\text{-th}$ iteration. The initial values are $C^0_1=1$ and $C^0_2=0$. The noise wave parameters ($T_{\text{unc}}$, $T_{\cos}$, and $T_{\sin}$) also take an initial value of zero, and in subsequent iterations they are modeled as polynomials in frequency. The polynomial coefficients are computed through a least squares fit to Equation~(\ref{equation_main_identity}) evaluated using the measurements of the open and shorted cable  and the current values of $C_1$ and $C_2$. This process converges in three iterations, after which $C_1$ and $C_2$ are also modeled as polynomials in frequency.

To find the optimum number of terms in the polynomials, the calibration quantities are modeled with increasing number of terms until the RMS difference between the physical temperature and the calibrated temperature spectra of the calibrators themselves reaches a minimum. It is found that seven terms are needed to model each of the five calibration quantities.

Figure~\ref{figure_calibration_quantities} presents the derived calibration quantities, computed as just described, along with the reflection coefficient of the receiver input.

\begin{figure*}[t]
\centering
\includegraphics[width=1\textwidth]{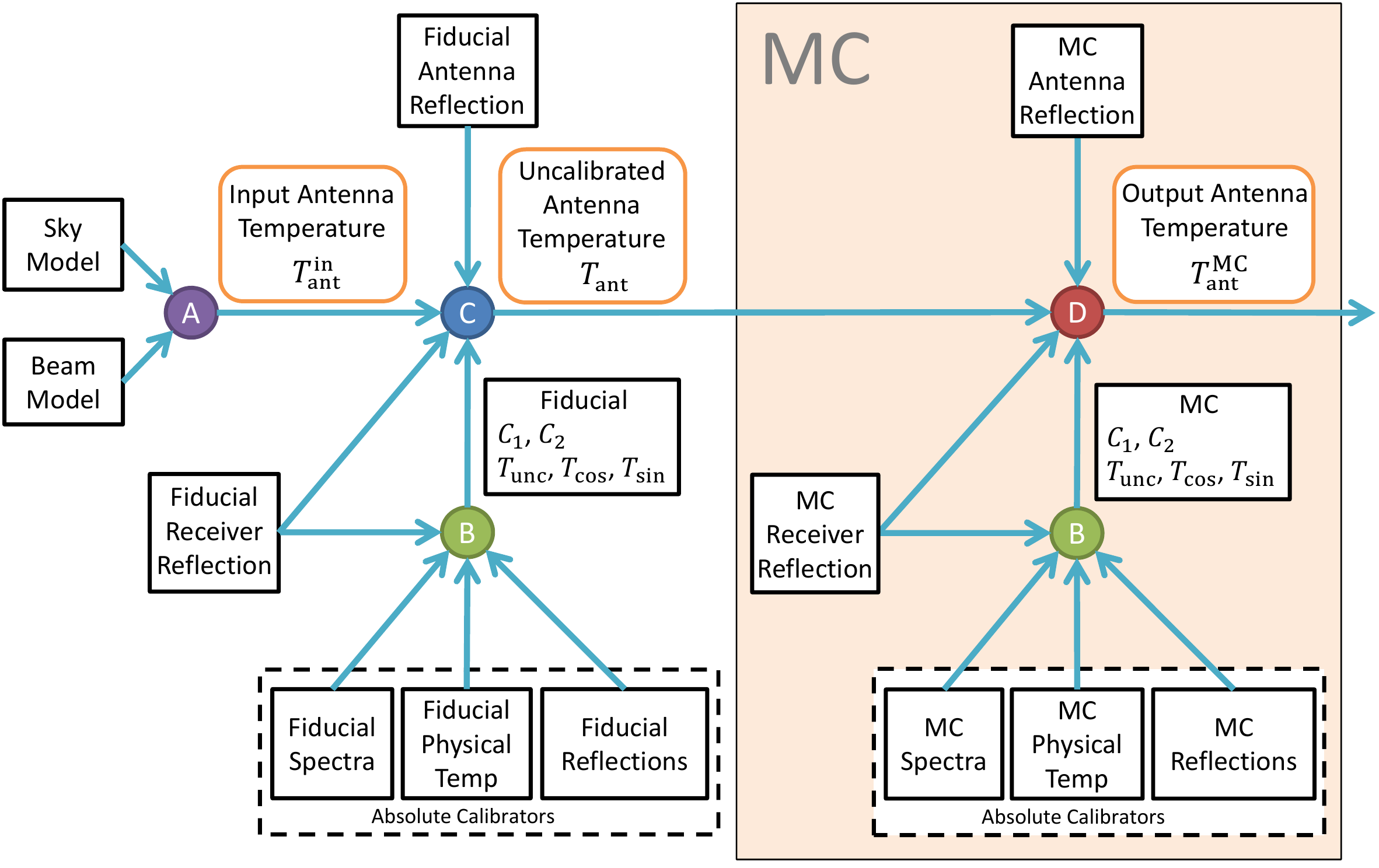}
\caption{Diagram of the Monte Carlo simulations used for propagating the receiver calibration uncertainties to the antenna temperature. The colored circles represent different operations: A) Equation~(\ref{equation_convolution}). B) Equation~(\ref{equation_main_identity}) solved for the calibration quantities. C) Equation~(\ref{equation_main_identity}) solved for $T^*_{\text{ant}}$. D) Equation~(\ref{equation_main_identity}) solved for $T_{\text{ant}}$. The blocks surrounded by dashed lines at the bottom represent the lab measurements of the four absolute calibrators. The MC section on the right-hand side represents operations conducted thousands of times for different realizations of the calibration measurements.}
\label{figure_MC_diagram}
\end{figure*}

\section{Propagation of Calibration Uncertainty}
\label{section_uncertainty_propagation}

Section~\ref{section_calibration_measurements} presented the computation of the receiver calibration quantities from laboratory measurements and introduced potential sources of uncertainty. In this section, we discuss these uncertainties in more detail and assess their impact on the final calibrated antenna temperature spectrum through a set of Monte Carlo (MC) simulations. We begin by outlining our MC uncertainty propagation pipeline and then motivate our choices for the measurement uncertainties used in it.

Figure~\ref{figure_MC_diagram} presents a diagram of our uncertainty propagation simulation scheme. From left to right, first, an ideal simulated input antenna temperature $T^{\text{in}}_{\text{ant}}$ is computed by convolving a sky model with an antenna beam model. Then, this antenna temperature is \emph{uncalibrated} by solving Equation~(\ref{equation_main_identity}) for $T^*_{\text{ant}}$ using the fiducial derived calibration quantities and receiver reflection coefficient from Section~\ref{section_calibration_measurements}, as well as a model for the reflection coefficient of the antenna.  Next, inside the block labeled MC, the antenna temperature is recalibrated with Equation~(\ref{equation_main_identity}) but using calibration quantities and reflection coefficients that potentially differ from their fiducial values due to simulated errors or noise estimated from our measurement uncertainties. This step is conducted for thousands of MC realizations of the perturbed calibration measurements. The output antenna temperatures are labeled $T^{\text{MC}}_{\text{ant}}$.   Finally, for each simulated output spectrum, we fit and subtract a polynomial to match the foreground subtraction procedure applied to actual measurements.  We use the resulting residuals to quantify the magnitude of the propagated calibration errors.

Our present interest is in understanding the role of the receiver in the experimental error budget.  Hence, we restrict our analysis here to only the effects of uncertainties from calibration of the receiver.  Our uncertainty propagation simulations assume perfect removal of antenna losses and beam chromaticity.  In addition, the simulations do not address potential changes in receiver performance during field operations due to, for instance, unaccounted temperature gradients or component aging. Field performance will be discussed in forthcoming papers that present the analysis of sky measurements. The rest of this section provides details of the simulations.

\subsection{Input Ideal Antenna Temperature}
\label{section_simulated_antenna_temperature}

The input antenna temperature is modeled by the convolution of a sky model with an antenna beam model:

\begin{equation}
T_{\text{ant}}^{\text{in}} = \frac{\int_{\Omega} T_{\text{sky}}\left(\theta,\phi\right)B\left(\theta,\phi\right) d\Omega}{\int_{\Omega} B\left(\theta,\phi\right)d\Omega},
\label{equation_convolution}
\end{equation}

\noindent where $T_{\text{sky}}$ is the sky model, $B$ is the beam model, $\theta$ and $\phi$ are the zenith and azimuth angles respectively, and $\Omega$ represents coordinates above the horizon.   Below $200$ MHz, the sky brightness temperature is dominated by foregrounds that are more than four orders of magnitude stronger than the expected cosmological signal \citep{haslam1982, rogers2008, mozdzen2016b}.  Since the dominant effects of receiver mis-calibration and uncertainty operate on the total sky signal, it is acceptable to neglect the cosmological $21$-cm signal in our sky model and only include the foreground contribution.  We model the foregrounds using the Global Sky Model (GSM) described in \citet{de_oliveira_costa2008}. 

For the antenna beam model, we use the simple frequency-independent azimuthally-symmetric expression used by \citet{pritchard2010}, 

\begin{equation}
B\left(\theta,\phi\right) = \cos^2\theta.
\label{equation_beam_model}
\end{equation}

\noindent Real beams are known to be more complex than this expression, introducing spectral structure into the antenna temperature that could limit the detection of the cosmological signal \citep{vedantham2014, bernardi2016, mozdzen2016a}. However, in these simulations it is assumed that the chromatic effects of the beam are perfectly known and removed.

\begin{figure}[t]
\centering
\includegraphics[width=0.48\textwidth]{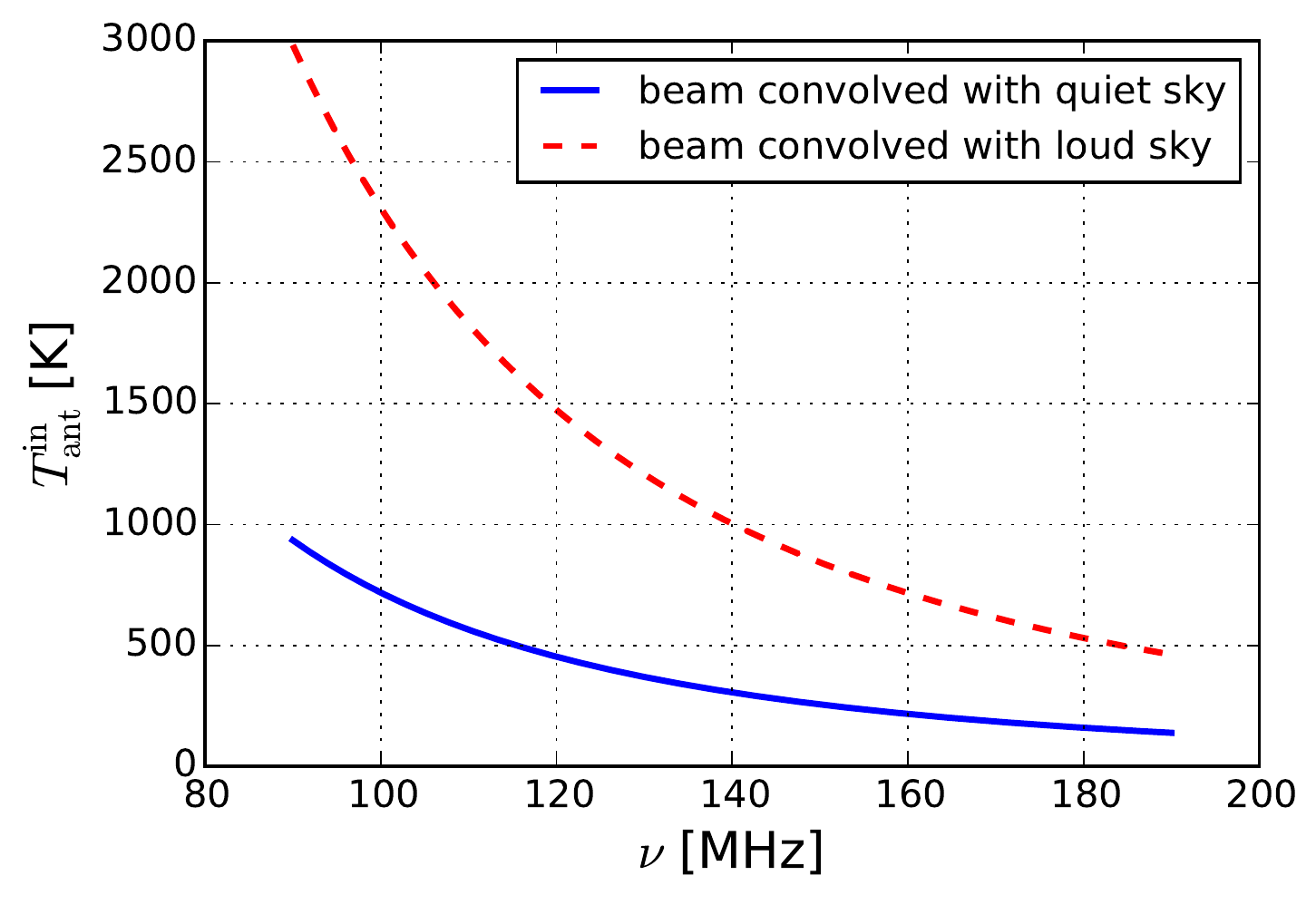}
\caption{Two antenna temperatures used as input in the simulations. They are computed by convolving the antenna beam model (Equation~(\ref{equation_beam_model})) with the foreground sky model (GSM, from \citet{de_oliveira_costa2008}). The quiet and loud skies correspond to high Galactic latitudes and the Galactic plane transit, respectively.}
\label{figure_antenna_temperature}
\end{figure}

The sensitivity of the instrument to errors in receiver calibration is a function of the antenna temperature. Ground-based instruments such as EDGES observe the sky, and thus foregrounds, continuously drifting over the antenna instead of conducting deep integrations on a single sky region. Therefore, the strength of the antenna temperature used for science analysis will vary as different parts of the sky drift through the beam. To account for the different foreground levels, we explore two scenarios  in the simulations: with the 1) lowest and 2) highest foreground contamination available at the EDGES observation latitude. These cases are labeled \emph{quiet} and \emph{loud} sky, respectively, and their convolution with the beam is shown in Figure~\ref{figure_antenna_temperature}.

\subsection{Antenna Reflection Coefficient}
\label{section_simulated_antenna_reflections}

Since we are mainly interested in understanding the effect of errors in receiver calibration, we assume a simple model for the antenna reflection coefficient.  Our model for the reflection magnitude is flat in frequency with a level of $-15$ dB. We model the phase as a linear decrease by $500^{\circ}$ between $90$ and $190$ MHz. This is realistic and corresponds to a delay of $500^{\circ}/(360^{\circ} \times 100\;\text{MHz})=13.88$ ns. This model reproduces, to first order, the measured EDGES antenna properties and is generally consistent with the performance achievable from a dipole-based antenna operating over an octave bandwidth.

\capstartfalse
\begin{deluxetable}{clcc}
\tabletypesize{\scriptsize}
\tablewidth{0pt}
\tablecaption{Nomenclature and Uncertainties of Receiver Calibration Measurements\label{table_nomenclature}}
\tablehead{\\ \colhead{Quantity} & \colhead{Device} & \colhead{$1\sigma$ Uncert} & \colhead{Note}}
\startdata\\[-1mm]
\multicolumn{2}{l}{\textbf{Uncalibrated Spectrum}}  &   & \\
\cline{1-2}\\[-1mm]  
$T^*_A$\;\dotfill     & Ambient Load  \dotfill    & 66 mK &\\[1mm]
$T^*_H$\;\dotfill     & Hot Load      \dotfill    & 66 mK & (i)\\[1mm]
$T^*_O$\;\dotfill     & Open Cable    \dotfill    & 95 mK &\\[1mm]
$T^*_S$\;\dotfill     & Shorted Cable \dotfill    & 95 mK &\\[1mm]
\\[1mm]
\multicolumn{2}{l}{\textbf{Physical Temperature}}   &   & \\
\cline{1-2}\\[-1mm]
$T_A$\;\dotfill                   & Ambient Load           &  &\\
$T_{Ht}$\;\dotfill                 & Hot Load Termination   & $100$ mK & (ii)\\
$T_O$\;\dotfill                    & Open Cable             &  &\\
$T_S$\;\dotfill                    & Shorted Cable          &  &\\
\\[1mm]
\multicolumn{2}{l}{\textbf{Magnitude of Reflection Coefficient}}   &  & \\
\cline{1-2}\\[-1mm]
$|\Gamma_A|$\;\dotfill                     & Ambient Load             &  &\\
$|\Gamma_H|$\;\dotfill                     & Hot Load                 &  &\\
$|\Gamma_O|$\;\dotfill                     & Open Cable               & 0.0001 & (iii)\\
$|\Gamma_S|$\;\dotfill                     & Shorted Cable            &  &\\
$|\Gamma_{\text{rec}}|$\;\dotfill          & Receiver                 &  &\\
$|\Gamma_{\text{ant}}|$\;\dotfill          & Antenna\tablenotemark{$\star$}  &  &\\
\\[1mm]
\multicolumn{2}{l}{\textbf{Phase of Reflection Coefficient}}  &  & \\
\cline{1-2}\\[-1mm]
$\angle\;\Gamma_A$\;\dotfill               & Ambient Load      &  &\\
$\angle\;\Gamma_H$\;\dotfill               & Hot Load          &  &\\
$\angle\;\Gamma_O$\;\dotfill               & Open Cable        &  0.015$^{\circ}/|\Gamma|$ & (iv)\\
$\angle\;\Gamma_S$\;\dotfill               & Shorted Cable     &  &\\
$\angle\;\Gamma_{\text{rec}}$\;\dotfill    & Receiver          &  &\\
$\angle\;\Gamma_{\text{ant}}$\;\dotfill    & Antenna\tablenotemark{$\star$}  &  &\\
\\[1mm]
\multicolumn{2}{l}{\textbf{Magnitude of Transmission Coefficient}}   &  & \\
\cline{1-2}\\[-1mm]
$|S_{21}|$\;\dotfill                        & Semi-rigid cable    & 0.015 & (v) \\
\enddata
\tablecomments{\newline(i) Uncertainty from thermal noise.\newline (ii) Same uncertainty for all.\newline  (iii) Same uncertainty for all, in linear scale.\newline (iv) Uncertainty is a function of reflection magnitude.\newline (v) In linear scale.\newline ($\star$) Antenna reflection measurements are not part of receiver calibration but their uncertainty is also accounted for in this study.}
\end{deluxetable}

\subsection{Measurement Uncertainties}

Here, we describe the uncertainties assigned to each of the calibration measurements from Section~\ref{section_calibration_measurements}, as well as the antenna reflection coefficient. These uncertainties are applied inside the MC block of the uncertainty propagation, to create slightly perturbed realizations of the calibration parameters. A summary of the nomenclature and values is presented in Table~\ref{table_nomenclature}.

In the top block of Table~\ref{table_nomenclature}, we show the $1\sigma$ uncertainties associated with each of the uncalibrated spectra of the absolute calibration loads ($T^*_A$, $T^*_H$, $T^*_O$, $T^*_S$).  The uncertainties are set equal to the RMS of the residuals from polynomial or Fourier series fits to the actual measured spectra.  In all cases, the residuals are noise-like.  The residual RMS for the ambient and hot load measurements is $66$~mK, while for the open/shorted cable measurements it is $95$~mK.  For our MC realizations of the perturbed spectra, we begin with the models for the fiducial spectra and add to each frequency channel realizations of Gaussian noise, uncorrelated from channel to channel, drawn from the assigned uncertainties.

The second block of Table~\ref{table_nomenclature} shows the uncertainties associated with the physical temperatures ($T_A$, $T_{Ht}$, $T_O$, $T_S$) of the calibrator sources. We measure all the physical temperatures with thermistors and a resistance-to-temperature model. The accuracy of the setup is estimated to be $100$~mK, hence we use that value for the uncertainty of all physical temperature measurements.  For the MC realizations of the calibrator temperatures, we draw from Gaussian distributions centered at the fiducial values with a $1\sigma$ width of $100$ mK.

The third and fourth blocks of Table~\ref{table_nomenclature} summarize the uncertainties applied to the amplitudes and phases of reflection coefficients, respectively.  Our MC modeling propagates the reflection coefficient uncertainties for: 1) the absolute calibration loads ($\Gamma_A$, $\Gamma_H$, $\Gamma_O$, $\Gamma_S$), 2) the receiver input ($\Gamma_{\text{rec}}$), and 3) the antenna ($\Gamma_{\text{ant}}$).  Although the antenna reflection coefficient is not measured as part of the receiver calibration, it is tightly coupled to final receiver performance and we account for its uncertainty here.  We also consider the uncertainty (listed in the fifth block in Table~\ref{table_nomenclature}) on the transmission coefficient, $|S_{21}|$, of the semi-rigid cable inside the hot load because it is relevant in the computation of the effective noise temperature from the device (Equation~(\ref{equation_available_power_gain})).

The main uncertainties in reflection coefficient measurements arise from uncertainty in VNA calibration. VNA calibration involves measuring three reflection standards at the VNA measurement plane: an open standard, a short standard, and a $50$-$\Omega$ standard. As part of the EDGES efforts to increase the accuracy in calibration measurements, we improved upon the manufacturer-specified VNA tolerances through more accurate modeling of the $50$-$\Omega$ standard \citep{blackham2005, scott2005, ridler2006, monsalve2016}. With the improved model for the standard, the additive uncertainty in the magnitude of measured reflection coefficients was reduced to a $1\sigma$ linear voltage ratio of $10^{-4}$, which is equivalent to $0.005$ dB for a reference reflection of $-15$ dB. In addition to residual modeling errors, sub-dominant effects that contribute to this uncertainty include VNA drifts and imperfect connection repeatability. Thus, the uncertainties assumed in the MC simulation for the reflection magnitudes are modeled as an additive contribution, fully correlated across the band, with an amplitude drawn from a Gaussian with a $1\sigma$ width of $10^{-4}$.

\begin{figure}[t]
\centering
\includegraphics[width=0.48\textwidth]{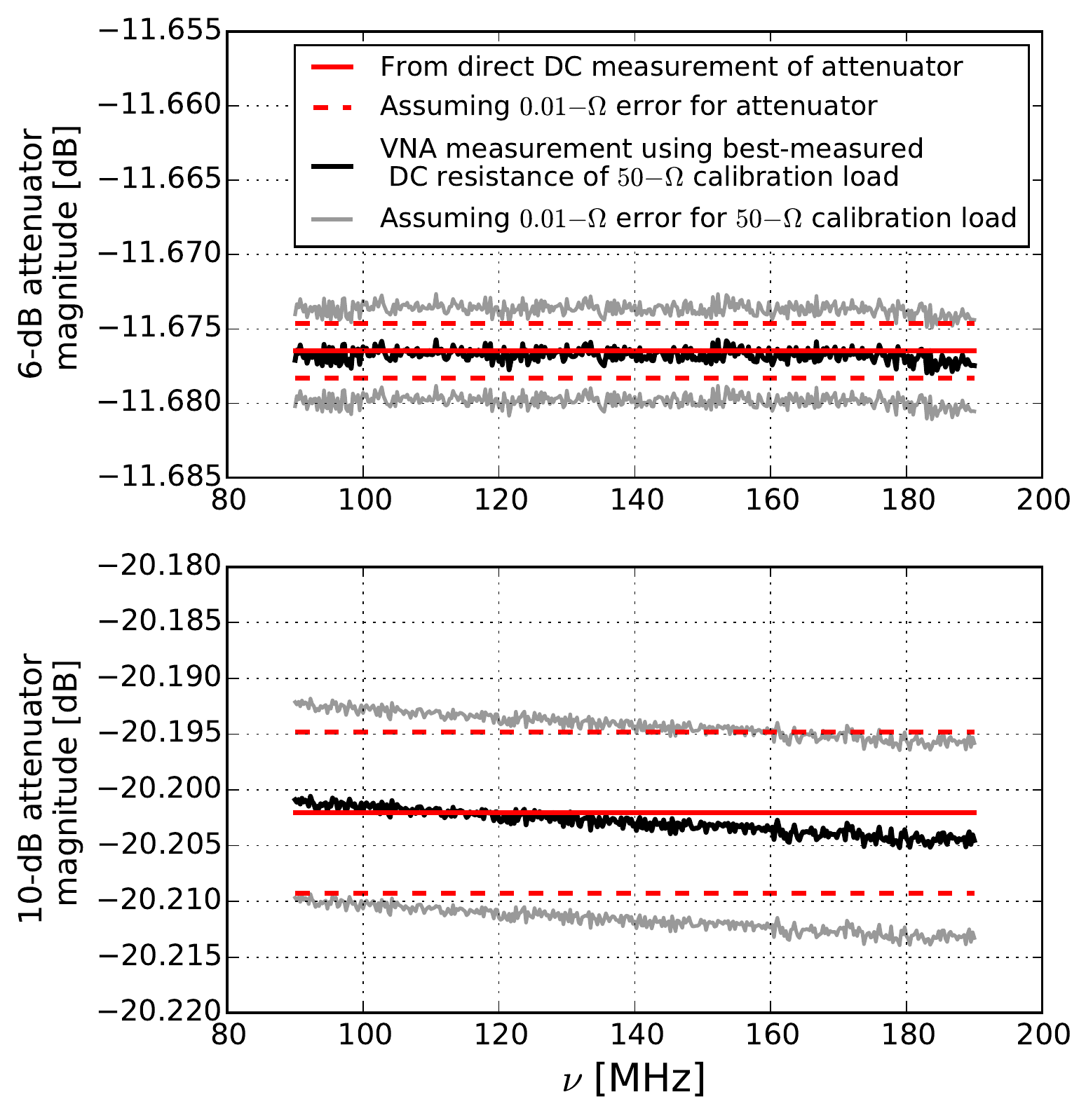}
\caption{Verification of accuracy in the magnitude of reflection coefficient measurements with a VNA. The verification consists of measuring the reflection of two open-ended RF attenuators ($6$- and $10$-dB respectively) after a high-accuracy VNA calibration (black line), and comparing these measurements with the expectations for the attenuators from their DC resistance (solid red line). The VNA calibration requires the DC resistance of the $50$-$\Omega$ calibration load. This resistance, as well as that of the attenuators, can be measured to better than $0.01$ $\Omega$. For reference, in the figure, the gray lines represent a pessimistic $\pm0.01$-$\Omega$ error in the calibration load resistance and the dashed red lines represents a $\pm0.01$-$\Omega$ error in the attenuator resistance. Even when assuming this pessimistic resistance error, the accuracy is better than $\pm 0.01$ dB at a reflection level of $\approx -20.2$ dB, and better than $\pm 0.005$ dB at $\approx -11.7$ dB.}
\label{figure_reflection_accuracy}
\end{figure}

Figure~\ref{figure_reflection_accuracy} presents a verification of the magnitude accuracy of our VNA measurements using open-ended RF attenuators.  In the figure, we show reflection coefficient measurements of an attenuator compared to  forecasts based on a direct DC measurement of the attenuator resistance.  In particular, for $-20$-dB reflection, the direct agreement is better than $0.003$~dB and remains better than $\pm0.01$~dB even if we assume a pessimistic error in the resistance measurement.

We model reflection phase uncertainty using the form $k/|\Gamma|$ found in typical VNA specifications.  Here, $k$ is a constant that represents the uncertainty value for the special case of a total reflection ($|\Gamma=1|$). This model captures the fact that it is more difficult to determine the phase for smaller reflections than for larger reflections. In our simulations, we use $k=0.015^{\circ}$ and model phase uncertainty as an additive contribution with a $1\sigma$ width given by $k/|\Gamma|$. For reference, this corresponds to a $\pm3\sigma$ range of $\pm0.25^{\circ}$ at the $-15$-dB magnitude of our antenna reflection coefficient model. As was the case for the reflection magnitude, the additive contribution to the phase is modeled as fully correlated across the band. This is a realistic first-order approximation for our VNA measurements of $100$-MHz bandwidth.

The uncertainty of transmission coefficient ($|S_{21}|$) measurents also benefits from the improvements in the measurement of reflection coefficient.  We assign $1\sigma$ uncertainty of $0.015$ to the transmission coefficient of the semi-rigid cable in the hot load calibrator and model it in the same way as the reflection magnitudes.

\begin{deluxetable}{crrrrrrrr}
\tablecolumns{9}
\tabletypesize{\scriptsize}
\tablewidth{0pt}
\tablecaption{$95\%$ Confidence RMS Residuals in Units of mK \label{table_residuals}}
\tablehead{ \colhead{} & \multicolumn{8}{c}{Number of Terms} \\
\cline{2-9} \\ 
\colhead{Source} & \colhead{0} & \colhead{1} & \colhead{2} & \colhead{3} & \colhead{4} & \colhead{5} & \colhead{6} & \colhead{7} }
\startdata
\\
$T^*_A$  & 5 &  4 &  4 &  4 &  3 &  3 & 2 & 2 \\
         & 25 & 22 & 20 & 18 & 15 & 12 & 6 & 1 \\
\\
$T^*_H$  & 5 &  5 &  4 &  4 &  3 &  2 & 1 & 1 \\
         & 23 & 22 & 20 & 19 & 17 & 14 & 9 & 2 \\ 
\tableline
\\
$T_A$    & 381  & 357 &  56 &  1 & & & & \\
         & 1852 & 340 &  63 & & & & & \\ 
\\
$T_{Ht}$ & 398  & 258 & 40  &  1 & & & & \\
         & 1958 & 240 & 47  &  1 & & & & \\
\\
$T_O$    & 1 & 1 & 1 & 1 & 1 & & & \\
         & 2 & 1 & 1 & 1 & 1 & & & \\
\\
$T_S$    & 1 & 1 & 1 & 1 &   & & & \\
         & 1 & 1 & 1 & 1 & 1 & & & \\
\tableline
\\
$|\Gamma_A|$       & 19 & 19 &  7 & 5 & 1 & & & \\
                   & 94 & 27 & 26 & 4 & 2 & & & \\
\\
$\angle\;\Gamma_A$ & 60 &  43 & 41 &  9 & 3 & 1 & & \\
                   & 228 & 136 & 27 & 22 & 2 & 1 & & \\
\\
$|\Gamma_H|$       & 22 & 16 &  9 & 5 & 1 & & & \\
                   & 136 & 42 & 36 & 3 & 2 & & & \\
\\
$\angle\;\Gamma_H$ & 74 &  38 & 35 & 10 & 2 & 1 & & \\
                   & 320 & 198 & 27 & 26 & 2 & 1 & & \\
\\
$|\Gamma_O|$       & 5 & 4 & 4 & 2 & 1 & 1 & 1 & 1 \\
                   & 5 & 3 & 3 & 2 & 1 & 1 & 1 & 1 \\
\\
$\angle\;\Gamma_O$ & 2 & 2 & 2 & 2 & 1 & 1 & 1 & \\
                   & 2 & 2 & 2 & 2 & 1 & 1 & 1 & \\
\\
$|\Gamma_S|$       & 4 & 4 & 3 & 2 & 2 & 1 & 1 & 1 \\
                   & 4 & 4 & 3 & 2 & 2 & 1 & 1 & 1 \\
\\
$\angle\;\Gamma_S$ & 1 & 1 & 1 & 1 & 1 & 1 & 1 & \\
                   & 1 & 1 & 1 & 1 & 1 & 1 & 1 & \\
\\
$|\Gamma_{\text{rec}}|$       & 15 & 15 & 15 & 14 & 10 &  5 &  5 & 1 \\
                              & 58 & 57 & 57 & 56 & 42 & 23 & 19 & 5 \\
\\
$\angle\;\Gamma_{\text{rec}}$ & 47 &  41 &  35 &  27 &  25 &  20 &  6 &  6 \\
                              & 176 & 160 & 138 & 118 & 105 &  81 & 27 & 24 \\
\\
$|S_{21}|$  & 104 &  68 & 11 & & & & & \\
            & 520 &  63 & 13 & & & & & \\
\tableline
\\
$|\Gamma_{\text{ant}}|$    & 32 & 12 & 12 & 12 &  9 &  6 &  3 & 1 \\
                           & 112 & 33 & 33 & 32 & 26 & 15 & 10 & 3 \\
\\
$\angle\;\Gamma_{\text{ant}}$   & 36 &  33 & 29 & 28 & 23 & 16 &  7 &  4 \\
                                & 100 &  92 & 79 & 73 & 63 & 46 & 19 & 12 \\
\tableline
\\
All          & 549 & 443 &  94 &  42 &  36 &  26 & 12 &  8 \\
             & 2803 & 504 & 178 & 146 & 128 &  96 & 37 & 27
\enddata
\tablecomments{For each uncertainty source, the top and bottom rows correspond to results for the quiet and loud skies respectively.}
\end{deluxetable}

\subsection{Simulated Foreground Subtraction}

Final cosmological parameter estimation for EDGES and similar experiments is performed by simultaneously fitting a  signal model with a parametrized foreground model.   In this step, the foreground model consists of a low-order polynomial or set of basis functions.   These functions are able to fit the foreground and beam chromaticity structure in the measured spectra, but also tend to absorb some of the expected $21$-cm signal.   Calibration errors that are similar to foreground structures---i.e. those that exhibit large spectral coherence---will be associated with the foreground model terms during the final parameter estimation and will have little additional impact on the $21$-cm signal estimation.  Calibration errors that vary relatively rapidly in frequency, on the other hand, will not be fit by the foreground model and will yield more interference with the signal estimation.  In order to take this effect into account in our analysis here, we  characterize the output antenna temperatures from our uncertainty propagation after fitting and removing a foreground model. We use the polynomial model given by:

\begin{equation}
\text{model} =  \sum\limits_{i=0}^{N-1} a_i \nu^{-2.5+i}. \label{equation_model2}
\end{equation}

\noindent This model was introduced in \citet{mozdzen2016a} as an efficient expression for removing foreground and beam effects. In the context of this paper, it shares the flexibility of a generic polynomial but is more efficient for residuals with a predominant $\beta\approx -2.5$ power-law behavior. It is chosen as a model that could help remove the foreground and residuals from calibration simultaneously with few terms.

\section{Results}
\label{section_results}

In this section, we report the results of our uncertainty propagation. The effect of each calibration uncertainty source is first analyzed individually by holding all other sources of uncertainties to zero. Then, all the uncertainties are propagated simultaneously to examine their combined impact.  

We characterize the significance of the calibration uncertainties by calculating the RMS of residuals to the foreground model fit for differing numbers of polynomial terms in the foreground model, exploring between zero and seven terms.  Distributions of RMS values are produced from thousands of MC repetitions.  We use $5000$ repetitions when a single effect is studied in isolation and $10^5$ when all the effects are considered simultaneously.  These RMS distributions are characterized in terms of their $95\%$ upper bounds, RMS$^{95\%}$.

\begin{figure*}
\centering
\includegraphics[width=1\textwidth]{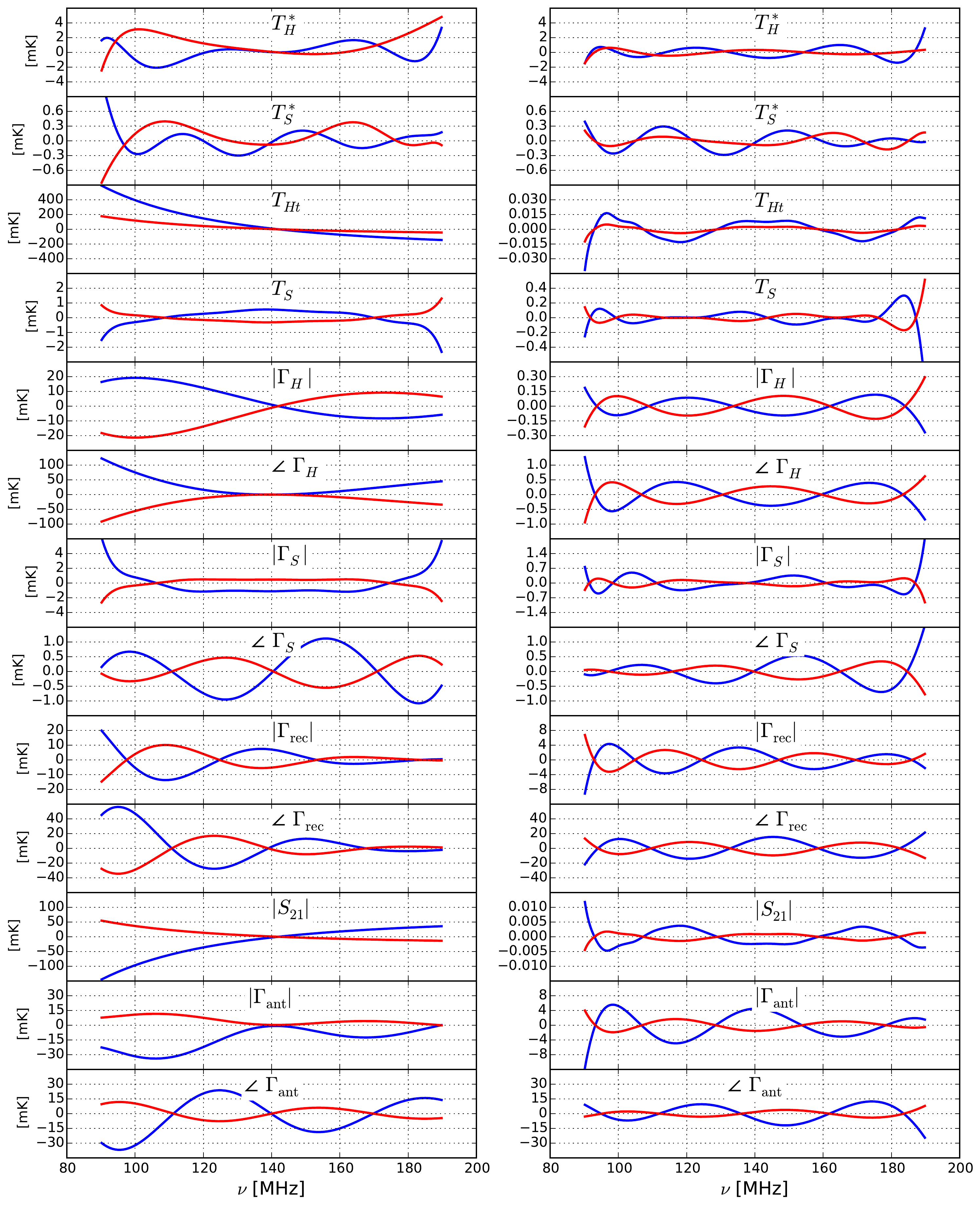}
\caption{\emph{Left}: Two examples (blue and red) of spectral distortions in the antenna temperature resulting from different calibration errors (different rows) for a quiet sky. \emph{Right}: Residuals for the same two cases after removing Equation~(\ref{equation_model2}) with five terms. From top to bottom, residuals are presented for the spectra, physical temperature, and reflection coefficient of the hot load and shorted cable (eight top rows). Also shown are results for the reflection coefficient of the receiver, the transmission coefficient magnitude of the semi-rigid cable inside the ambient/hot load, and the reflection coefficient of the antenna. Results for the ambient load and open cable are not shown because they are similar to those for the hot load and shorted cable. }
\label{figure_residuals}
\end{figure*}

The results are presented in Table~\ref{table_residuals}. The table lists the uncertainty sources and the corresponding $\text{RMS}^{95\%}$ levels after removing the foreground model, with increasing numbers of polynomial terms shown in separate columns from left to right. The two rows next to each source represent the residuals for the quiet (top) and loud (bottom) skies respectively. For residuals below $1$~mK, the cells are left empty. In the case of the the open and shorted cable spectra ($T^*_O$, $T^*_S$) the residuals are always below $1$ mK and hence not shown. Figure~\ref{figure_residuals} illustrates typical residuals for a quiet sky. Each row corresponds to a different source of uncertainty. The left column shows two representative cases (in blue and red) drawn arbitrarily from our MC simulations before removing any term. The right column shows the residuals for the same two cases after fitting and removing Equation~(\ref{equation_model2}) with five terms.

\subsection{Interpretation}

Several trends can be identified from Table~\ref{table_residuals}. First and as expected, for most error sources and polynomial orders the residuals are larger when observing a loud sky. This motivates the preference of low-foreground observations for estimation of the cosmological signal, or the downweighting of strong-foreground regions as suggested, for instance, in \citet{liu2013}.

The table also makes evident the low sensitivity of the calibration to errors in measurements of the open and shorted cable. As stated before, the residuals for their spectra ($T^*_O$, $T^*_S$) remain consistently below $1$ mK and are not shown in the table. For their physical temperatures ($T_O$, $T_S$) the initial residuals are $\leq 2$ mK and for the reflections ($\Gamma_O$, $\Gamma_S$) they are $\leq 5$ mK. At five terms they all reach $1$ mK or less.

The largest initial residuals occur for errors in the physical temperatures of the ambient and hot loads ($T_A$, $T_{Ht}$) which, for the loud sky, reach up to $\approx 2$ K. However, as shown in Figure~\ref{figure_residuals}, they strongly follow a power law and can be removed with few terms. Specifically, three terms are needed to reduce the residuals to $\leq 1$ mK for our foreground model. The same applies to the transmission coefficient magnitude of the cable inside the ambient/hot load ($|S_{21}|$) since this quantity is involved in the computation of the hot load noise temperature (Section~\ref{section_absolute_hot_load}).

The initial impact of thermal noise in the ambient and hot spectra ($T^*_A$, $T^*_H$) is at the $5$-mK and $\approx25$-mK level for the quiet and loud skies respectively. These residuals decrease slowly as terms are added. More than five terms would be needed to reach $1$ mK. For the reflections of these loads ($\Gamma_A$, $\Gamma_H$) the initial residuals are significantly higher (up to $320$ mK). However, five terms are sufficient to reduce them to $\leq 1$ mK due to their smooth spectral shape.

For the receiver and antenna reflection coefficients ($\Gamma_{\text{rec}}$, $\Gamma_{\text{ant}}$) the initial residuals are comparable to those for the ambient and hot loads. However, it takes more than seven terms to reduce them to $\leq 1$ mK. With three or more terms removed these residuals are the largest among all the effects. The similarity in behavior between the two sources is not surprising considering their strong interaction in the calibration equations (Section~\ref{section_calibration_formalism}).

Finally, the last rows of the table present the residuals when all the uncertainties are combined in a comprehensive Monte Carlo analysis. As expected, initially they are larger than those from individual sources, amounting to about $0.5$ and $2.8$ K for the quiet and loud skies respectively. For the quiet sky they drop to $26$ mK with five terms.

In the full uncertainty propagation, errors in reflection coefficients are assumed uncorrelated. However, a test was also conducted with correlated errors where, in any given MC repetition, the same error was applied to all the magnitudes/phases. This exercise simulates a case where all these measurements are affected by the same VNA bias. The results are qualitatively and quantitatively very similar to those shown in the table, thus not shown for brevity.

\section{Reduction of Systematics}
\label{section_discussion}

As shown in \citet{pritchard2010} and \citet{morandi2012} through simulations, over a frequency range similar to EDGES High-Band and with sub mK channel noise, it is possible to probe a significant range of EoR models when the foreground and instrument spectral response are simultaneously modeled with less than six polynomial terms. Some recent cosmological models yield large or late absorption features, or rapid reionization histories within the EDGES High-Band range \citep{mirocha2013, fialkov2014a, fialkov2014b, kaurov2015, fialkov2016a, fialkov2016b, mirocha2016, cohen2016}. These models should also be accessible with a similar instrument performance.

The EDGES beam chromaticity has been modeled by \citet{mozdzen2016a} to be less than $1$ mK at quiet sky locations after a fit to Equation~(\ref{equation_model2}) with five terms. Ideally, this number of terms and residual level should not increase due to systematics from receiver calibration. Therefore, in order to avoid being limited by receiver systematics, the natural goal for the receiver calibration performance corresponds to keeping the residuals below $1$ mK after a fit to Equation~(\ref{equation_model2}) with five terms.

This goal is already being met for most of the measurements. The three exceptions correspond to 1) the spectra of the ambient and hot loads, 2) the reflection coefficient of the receiver, and 3) the reflection coefficient of the antenna. Simulations indicate that reducing the impact of noise in the ambient and hot load spectra to $\leq 1$ mK with five terms requires lowering the noise level to $\lesssim 40$\% of its nominal value. With the current digitization efficiency, this is equivalent to integrating for more than $0.40^{-2}=6.25$ days, since the nominal value is the result of one day of integration. Meeting this requirement is realistic and planned for future calibration revisions.

The situation is more challenging for the reflection coefficients. It would be necessary to reduce the magnitude and phase uncertainties to $\lesssim 20$\% and $\lesssim 7$\% of the nominal values, respectively. Achieving such tight tolerances is an area of active research and we are identifying promising techniques. For instance, the current uncertainty levels could be reduced by incorporating measurements of custom reflection standards, such as well characterized air-dielectric coaxial lines, into the VNA calibration process \citep{eio2006, wubbeler2009, robertsmartens2014}.

\subsection{Improvement of Impedance Match}

\begin{deluxetable}{cccccccccc}
\tabletypesize{\scriptsize}
\tablewidth{0pt}
\tablecaption{$95$\% Confidence RMS Residuals for a Quiet Sky due to Errors in Antenna Reflection Coefficient, in mK\label{table_antenna_residuals}}
\tablehead{\colhead{} & \colhead{} & \multicolumn{8}{c}{Number of Terms} \\
\cline{3-10} \\ 
\colhead{Source} & \colhead{Case} & \colhead{0} & \colhead{1} & \colhead{2} & \colhead{3} & \colhead{4} & \colhead{5} & \colhead{6} & \colhead{7} }
\startdata
\\
$|\Gamma_{\text{ant}}|$        & (a) & 32 & 12 & 12 & 12 &  9 & 6 & 3 & 1 \\
                               & (b) & 29 &  7 &  6 &  6 &  5 & 3 & 1 & 1 \\
                               & (c) & 21 & 12 & 12 & 11 &  9 & 6 & 3 & 1 \\
                               & (d) & 34 & 12 & 11 &  6 &  3 & 1 &   &   \\
                               & (e) & 18 &  6 &  6 &  6 &  5 & 3 & 1 & 1 \\
                               & (f) & 19 &  7 &  6 &  3 &  2 &   &   &   \\  
\\
$\angle\;\Gamma_{\text{ant}}$  & (a) & 36 & 33 & 29 & 28 & 23 & 16 & 7 & 4 \\
                               & (b) & 18 & 17 & 16 & 15 & 13 &  8 & 4 & 1 \\
                               & (c) & 35 & 32 & 29 & 27 & 23 & 15 & 7 & 4 \\
                               & (d) & 36 & 33 & 25 & 16 &  6 &  3 &   &   \\
                               & (e) & 18 & 17 & 15 & 15 & 12 &  7 & 4 & 1 \\
                               & (f) & 18 & 17 & 12 &  9 &  2 &  1 &   &

\enddata
\tablecomments{\newline Case (a): Nominal. Same as in Table~\ref{table_residuals}.\newline Case (b): After lowering $|\Gamma_{\text{rec}}|$ from $\approx -20$ dB to $-30$ dB.\newline Case (c): After lowering $|\Gamma_{\text{ant}}|$ from $-15$ dB to $-20$ dB.\newline Case (d): After changing antenna delay from $13.88$ ns to $6.94$ ns. \newline Case (e): Combined (b) and (c). \newline Case (f): Combined (b), (c), and (d).}
\end{deluxetable}

An alternative approach to reduce the sensitivity of EDGES to errors in reflection measurements consists of improving the impedance match between the antenna and receiver. As stated in Section~\ref{section_calibration_formalism}, our reflection coefficients are referenced to the $50$-$\Omega$ system impedance. Therefore, improvement involves bringing both impedances closer to $50$-$\Omega$ or, equivalently, lowering both reflection coefficients.

Table~\ref{table_antenna_residuals} presents five simulation examples with potential for reducing sensitivity to errors in the antenna reflection coefficient (magnitude, $|\Gamma_{\text{ant}}|$, and phase, $\angle\;\Gamma_{\text{ant}}$) for different levels of reflection from the antenna and receiver. The computations assume a quiet sky and the nominal uncertainties of Table~\ref{table_nomenclature}.

Case (a) corresponds to the same residuals shown in Table~\ref{table_residuals}, presented again for reference. In case (b), the magnitude of the receiver reflection coefficient is lowered from $\approx -20$ to $-30$ dB. This change produces a reduction in residuals from the antenna reflection magnitude and phase by a factor of about two for almost any number of terms. In case (c), the nominal settings have been modified by lowering the magnitude of the antenna reflection coefficient from $-15$ to $-20$ dB. Relative to (a), this scenario produces an improvement in the initial residuals of the magnitude, which decreases from $32$ to $21$ mK. However, when introducing terms in the model, the residuals become almost identical to those in the nominal case. In case (d) the delay of the antenna has been reduced from $13.88$ ns to $6.94$ ns, i.e., the linear change in phase between $90$ and $190$ MHz has been reduced from $500^{\circ}$ to $250^{\circ}$. This change has no impact on the initial residuals, but as more terms are introduced the improvement becomes significant. With five terms, the magnitude residuals are reduced from $6$ to $1$ mK, and in phase they are reduced from $16$ to $3$ mK. Case (e) corresponds to the combination of cases (b) and (c), i.e., both reflection magnitudes have been reduced. Case (f) combines all the suggested improvements and reduces the residuals to the lowest figures. They start at 19 and 18 mK for the magnitude and phase with no terms removed, and converge to $\leq 1$ mK with five terms.

The improvements described above have the potential to reduce sensitivity to errors. At the same time, their implementation has to consider tradeoffs typically encountered in wideband instrument design. For instance, reducing the antenna reflection coefficient can have an impact on the spectral smoothness of the beam. Similarly for the receiver, it is necessary to balance its input reflection coefficient with its noise performance. In addition, the phase of the antenna reflection depends on the physical dimensions of the antenna and on the length of any transmission line between the antenna and the reflection measurement plane. We are currently investigating these alternatives to converge to an optimal instrumental solution.

\subsection{Reduction of Bandwidth}

The results presented thus far correspond to $95$\% residuals in the range $90-190$ MHz. On top of refinements in the instrumentation, a decrease in residuals could be achieved by reducing the bandwidth used in the science analysis motivated, for instance, by the interest in probing late ($z_r \lesssim 10$) reionization transitions \citep{robertson2015, planck2016}.

Here we show the improvement in performance when reducing the bandwidth to $80$ MHz by computing residuals in the range $110-190$ MHz ($11.9 \gtrsim z \gtrsim 6.5$) from the MC simulations of Section~\ref{section_uncertainty_propagation}. Discarding a $20$-MHz section at the low-frequency end results in less spectral structure, and in structure with lower amplitude due to the lower sky temperature in the remaining band. As depicted in Figure~\ref{figure_residuals}, this especially occurs for key sources of uncertainty such as the antenna and receiver reflection coefficients.

With a quiet sky and after removing five terms in the range $110-190$ MHz, the residuals for the ambient and hot load spectra are reduced to $1$ mK. For the magnitude and phase of the receiver reflection coefficient they go down to $3$ and $5$ mK, respectively. For the antenna reflection they are reduced to $2$ and $8$ mK. Finally, when all the uncertainty sources are considered the result is $9$ mK. This result for the combined case is about three times better than the $26$ mK obtained for the $90-190$ MHz range. This suggests that for a given signal model it may be possible to optimize the bandwidth processed to maximize the constraints on the model.

\section{Conclusion}
\label{section_conclusion}

This paper describes the laboratory calibration of the EDGES High-Band receiver used in the $2015-2016$ observational campaign. The calibration was performed prior to instrument deployment and involved measuring the spectra, reflection coefficients, and physical temperatures of four absolute calibrators connected at the receiver input. These measurements are used to determine the function that converts the antenna noise power to noise temperature. The calibration was done in the range $90-190$ MHz, corresponding to $14.8 \gtrsim z \gtrsim 6.5$.

To evaluate the impact on science analysis of realistic calibration uncertainties, we developed a Monte Carlo uncertainty propagation pipeline. The results are presented as $95$\% confidence levels for RMS residuals after modeling and fitting the systematics from simulated calibration errors with low-order polynomials, using between zero and seven terms.  We focus on the results from the five-term model because this is the number of terms required to account for beam chromaticity from the EDGES blade antenna.  After polynomial subtraction, the systematics from a combination of all the error sources considered in this work amount to $26$ mK for observations of a quiet, low-foreground sky. To a large extent, this value is the result of uncertainties in the reflection coefficient of the antenna and the receiver input, which contribute individually with $\lesssim 20$ mK. For almost all other error sources the residuals are $\leq 1$ mK.

The receiver calibration is determined in the lab and applied to data taken in the field. While every effort is made to equalize the instrument characteristics in the two circumstances, the validity of the assumptions that go into the MC simulations can only be tested with real sky measurements. This will be addressed in forthcoming papers.

Assuming that the receiver performance determined in the lab adequately reflects the instrument performance in the field, it should be possible for EDGES High-Band to probe some $21$-cm models described in the literature, particularly those with astrophysical parameter combinations that yield large or late absorption features, or with rapid reionization histories. Astrophysical results will also be reported in future papers.

There are promising near-term opportunities for reduction of systematics from reflection measurements: 1) the incorporation of custom reflection standards into the calibration of the vector network analyzer used for measurements, 2) the improvement of the impedance match between the antenna and the receiver input, 3) the reduction of the changes in antenna phase with frequency. Current efforts are focused on investigating and implementing these improvements. 

We thank the referee for suggestions that helped improve the quality of this manuscript. This work was supported by the NSF through research awards for the Experiment to Detect the Global EoR Signature (AST-0905990 and AST-1207761) and by NASA through Cooperative Agreements for the Lunar University Network for Astrophysics (NNA09DB30A) and the Nancy Grace Roman Technology Fellowship (NNX12AI17G). Raul Monsalve acknowledges support from the NASA Ames Research Center (NNX16AF59G).

\acknowledgments

\clearpage
\end{document}